\documentstyle[aaspp4]{article}

\def\lii{{\ion{Li}{1}}}

\def\lte{{\sc lte}}
\def\nlte{{\sc nlte}}
\slugcomment{Accepted by ApJ Letters}

\lefthead{Kiselman}
\righthead{Lithium lines in granulation}

\begin{document}



\title{Formation of \lii\ lines in photospheric granulation}

\author{Dan Kiselman}
\affil{The Royal Swedish Academy of Sciences, Stockholm Observatory,
       SE-133~36 {\sc Saltsj\"obaden}, Sweden; dan@astro.su.se}

\begin{abstract}
The possibility of significant systematic errors due to the use of 1D
homogeneous atmospheres in lithium-abundance
determinations of cool stars motivates a study of
non-local-thermodynamic-equilibrium (\nlte) effects on \lii\ line
formation in a 3D solar-granulation simulation snapshot. The
\nlte\ effect on the equivalent
width of the 671\,nm resonance line is small in 1D models or in integrated
light from the granulation model. The line-strength variations over
the granulation pattern are
however markedly different in \nlte\ compared to \lte\ -- observations of this
may provide diagnostics to \nlte\ effects. The effects of horizontal photon
exchange found in the granulation model are
moderate and due entirely to bound-bound processes, ultraviolet overionization
is unimportant.

\end{abstract}


\section{Introduction}
Stellar lithium abundances are potentially very useful for testing
astrophysical theories. The abundances
derived from observed spectra have spawned a number of scientific
debates, for a recent review see Thorburn (1996), also the conference proceedings
of Crane (1995) and Spite \& Pallavicini (1995), and the
introduction to Carlsson et al. (1994). It is important in this
context that we can be confident in the derived abundances -- which so far
mostly have been derived using the questionable assumptions of line
formation in local thermodynamic equilibrium (\lte) and plane-parallel
homogeneous \lte\ photospheres.

Efficient computer codes and extensive atomic data sets have made
possible realistic \nlte\ spectral-line modeling for light atoms in
plane-parallel cool-star
photospheres that can provide \nlte\ abundance corrections for the
convenience of the stellar-abundance community (e.g., Carlsson et
al. 1994; Kiselman \& Carlsson 1996).

A notable departure from plane-parallel homogeneity in the quiet solar
photosphere is the granulation, now understood as
a visible manifestation of convection
below the photosphere (e.g., Spruit, Nordlund, \& Title 1990). We
have also some knowledge of granulation on stars
adjacent to the Sun in the HR diagram (Gray \& Nagel 1989; Nordlund \&
Dravins 1990; Dravins \& Nordlund 1990a, 1990b).
How does granulation influence line strengths and abundance determinations?
Holweger, Heise, \& Kock (1990) argued that
solar abundance ratios should not be seriously in error since they are
based on line-strength ratios which are approximately constant over the
solar granulation pattern in the simulations of
Steffen (1989). This notion was confirmed observationally by
Kiselman (1994a) who found that lines of both neutral
and singly ionized species of several elements behaved
similarly by being stronger in bright granular
regions and weaker in dark intergranular lanes.
For other work on lines in realistic granulation simulations that is
relevant for abundance analysis, see Nordlund (1984), Bruls \&
Rutten (1992), Atroshchenko \& Gadun (1994), and Kiselman \& Nordlund
(1995).

Of special interest is the question of \nlte\ effects on line
formation in granulation.
Nordlund (1984) found that 3D \nlte\ effects could be of some
importance for \ion{Fe}{1} lines in solar granulation, while Kiselman
\& Nordlund (1995) found rather small such effects for \ion{O}{1}.
Note also that Mihalas, Auer, \& Mihalas (1978) did not find any
important 2D effects in their investigation of artificial but
solar-like atmospheric structures.
Kurucz (1995) claimed that \nlte\ effects in extremely metal-poor solar-type
stars can cause standard analyses to underestimate lithium
abundances with a factor of ten.
In this
scenario, \lii\ lines will be weak in hot photospheric regions where Li is
largely ionized. The cooler regions
would show strong \lii\ lines if \lte\ was valid, but the ultraviolet
radiation from adjacent hot regions will keep lithium largely ionized
also there. The \lii\ resonance line in integrated light would thus be very 
much weaker than the result from a 1D model representing a spatial
and temporal average of the photospheric structure. Thus the abundance
would be underestimated when an analysis is performed using standard
plane-parallel models. So far, there is no quantitative model to
verify this effect, which would certainly have important impact on the
debates related to stellar lithium abundances.

This Letter reports on \nlte\ aspects of \lii\ line
formation as found with experiments on a 3D solar-granulation
model. The results are thus directly applicable only for Sun-like
stars, but they should give some indication on what we can expect for
other stars as well.

\section{Simulations}

\subsection{Methods}
The \lii\ resonance doublet at
671\,nm is very weak in the solar spectrum since the solar
lithium abundance is so low -- the standard value used here
is $A_{\rm Li} = \lg {N({\rm Li})\over N({\rm H})} + 12 = 1.16$
(Grevesse, Noels, \& Sauval 1996; M{\"u}ller, Peytremann, \& de la
Reza 1975). The feature is blended
with several other weaker lines, the stronger (shorter wavelength)
doublet component being less affected.
The doublet splitting, together with hyperfine and isotopic splitting,
is neglected here since the weakness of the line
makes its equivalent width insensitive to line-broadening details.
Thus the lithium atomic model of Carlsson et
al. (1994) is employed with the change that the 671\,nm resonance
doublet is treated as a single line.
In principle, this treatment 
leads to a change in line formation height and some
modification of the NLTE behavior. This is, however, not a problem here --
the robustness of the results is indicated by the qualitative
behavior of the line being unchanged even when the abundance is increased
with a factor of ten.

The 3D photospheric model is a single snapshot from the
hydrodynamic granulation simulations of Stein \& Nordlund (1989) that
was also used by Kiselman \& Nordlund (1995) -- see that paper for
illustrations. It has been contracted to a $64\times
64 \times 55$ grid of thermodynamic quantities corresponding to
$6 \times 6 \times 1$\,Mm on the solar surface and showing more than
ten granules. (The figures of this paper only display $32\times 32$
surface points.) This single snapshot may not be adequate for statistically
reliable results, but it should show a representative sample of
different kinds of photospheric behavior.
  
Line profiles are calculated with a hybrid technique
using version 2.2 of the 1D \nlte\ code {\sc multi} (Carlsson 1986) 
which uses the operator perturbation techniques of Scharmer \&
Carlsson (1985).
The radiative bound-free transitions, and some bound-bound transitions in the
experiments below, are treated as fixed rates using a background
3D radiation field computed in (strong) \lte\ with the methods of
Kiselman \& Nordlund (1995). The equation
of transfer was solved along a set of rays -- five inclination 
angles ($\cos \theta = \mu$) and eight azimuthal angles ($\phi$)
-- and the resulting intensities along each ray were used to
calculate a mean intensity $J_\nu$ in each $(x,y,z)$ point.
The treatment can thus not really be called ``3D \nlte'' but it does
include 3D effects and \nlte\ effects.

All line results presented here are for disk center, i.e. for vertical rays
with $\mu = 1.0$.

\subsection{\lte\ and the 1.5D case}
The general diagnostic to be discussed here is the
dependence of line strength on continuum intensity.
The upper left panel of Fig.~\ref{fig_1.5D_33} shows equivalent width
of the 671\,nm line plotted as a function of continuum intensity when
the line is computed in \lte. Each point represents the vertically
emergent spectrum from one $(x,y)$ column of the simulation snapshot. Six points
marked with letters have been chosen as examples of different kinds of
regions. The neighboring plots show the local Planck
function (normalized with the emergent mean continuum intensity from the
snapshot) and the total opacity at line center (normalized on the
continuum opacity) at these selected locations. The Planck-function plot
naturally also illustrates the temperature structure. Note the
typical crossing over of temperatures, with the hot bright regions
(i.e. granules) like A and B having much steeper temperature gradients
than a typical intergranular region like F. C is an intermediate point
and D and E represents dark intergranular regions covered with cooler gas.

The over-all appearance of the \lte\ $I_c - W$ diagram is typical of
most lines studied so far (Kiselman 1994a). The strength of the \lii\
line generally increases with continuum brightness, but there is a significant
scatter, especially in the darker regions.
It is possible to understand this behavior in a simple qualitative
way. The line is strong where the temperature gradient is steep and/or if the
temperature in the line-forming layers is low. These very weak lines
are formed in a fairly extended region around $\lg \tau_c \approx
-1.5$, as evidenced from inspection of the contribution function due
to Magain (1986).

The large scatter in the \lte\ $I_c - W$ diagram
is due to the high temperature sensitivity of the line opacity and reflects
the variation in temperature in the upper layers of the granulation model.
Note how the dark examples (D, E, F) have similar low temperatures
where the continuum is formed -- the respective curves meet in
the upper right panel around $\lg \tau_c = 0$ -- and thus similar
continuum brightness. Higher up in the line-forming layers the curves
diverge, causing the line opacity to differ via the Saha ionization
equilibrium. Thus the difference in equivalent width ($W({\rm E})
> W({\rm D}) > W({\rm F})$) is a direct reflection of the
temperatures there with higher temperatures meaning less \lii\ and
weaker lines.

The plots in Fig.~\ref{fig_osmarcs_33} show results for an
effective-temperature range of plane-parallel photospheric models
computed with the {\sc osmarcs} code of Edvardsson et al. (1993). 
The \lte\ line behavior is much simpler in these models than in the
granulation snapshot.

Let us then consider the results when the lines are calculated
in \nlte\ and 1.5D geometry. This term is used here to signify
 that each vertical column in
the granulation snapshot is treated as a plane-parallel photosphere,
thus neglecting any horizontal photon exchange in the model. The lower left panel of 
Fig.~\ref{fig_1.5D_33} shows the resulting $I_c - W$ plot which differs
from the \lte\ one in three important respects.
First, the line is on the average weaker in \nlte.
Second, the \lte\ case shows increasing line strength
      in brighter regions while the opposite is true in \nlte.
Also striking is the narrow
      brightness dependence in the \nlte\ plot which contrasts to the
      big scatter among the points in the \lte\ plot. 

The \nlte-\lte\ difference in integrated-light equivalent widths
is about 30\,\%,
which is more than the 10\,\% found in the 1D results of
Fig.~\ref{fig_osmarcs_33}. Considering, however,
uncertainties in abundances, $f$-values, fundamental stellar
parameters, etc., such effects are barely significant in typical
stellar work. It is thus hardly possible to test either the \nlte\ or
the granulation results in integrated light alone.
The different appearance of the \lte\ and the \nlte\ plots shows,
however, that even if effects of departures from \lte\ look
insignificant in integrated light, or in 1D models, they can cause a
definite qualitative difference in the spatially resolved $I_c - W$
behavior. This supports the idea (Kiselman
1996) that \nlte\ effects can be discovered and diagnosed in this way.

The 1.5D \nlte\ plot is in fact easily understood. When we leave the
\lte\ approximation, the atomic level populations will be more or less
decoupled from the local kinetic temperature and the radiation field
will govern them instead. Thus the relevant quantities for the line,
its line source function and the line opacity, will be set by the
radiation field. This is exactly what the plot shows: for each value
of the background continuum intensity the line equivalent width is the
same.
The low lithium abundance will make all \lii\ lines and
ionization edges very weak, having virtually no impact on the
radiation field -- thus we only have to consider the continuum
background radiation fields. The weakness of the \lii\ 671\,nm line 
together with the fact that the 
line-photon-destruction probability is rather small ($\epsilon < 0.1$), 
makes the line source function $S^l$ follow
the continuum angular mean intensity $J_\nu$ rather closely.
A greater $J_\nu$ will thus lead to a greater $S^l$ and a weaker
line. Furthermore, the line opacity is set by the population of the
\lii\ ground state and so it is the ionization balance between \lii\
and Li\,{\sc ii} that sets the line opacity. In the \nlte\ case, the
radiation field (in the ionization edges themselves or at other
wavelengths) is obviously important for the ionization balance.

This explains why the range of 1D models in Fig.~\ref{fig_osmarcs_33}
seems a much better approximation to the granulation-simulation results in
\nlte\ than in \lte. The radiation temperature is set in the deeper
layers where the granulation structure is much more like the 1D models
than what is the case higher up.

\subsection{Diagnostic experiments}
The interest here is in finding the processes important for the
overall picture, not to explain \nlte\ mechanisms operating in individual
columns of the granulation snapshot.
The $I_c - W$ plots in Fig.~\ref{fig_icw22} illustrate what happens
when the 1.5D results already discussed are perturbed. These experiments
will be referenced to according to the figure's panel designations.

The importance of overionization caused by by $J_\nu >
B_\nu$ in bound-free edges is assessed by computing the photoionization
rates with $J_\nu = B_\nu$ ({\bf bf~B}). The small change shows that
ultraviolet overionization is unimportant.

A large majority of all \lii\ is in the ground state that sets the
671\,nm-line opacity. It will depart from its \lte\ value in a way dependent
of the excitation balance among the \lii\ levels since departures from
Boltzmann equilibrium  have an impact on the ionization
equilibrium. The general rule is that there will be overionization if
the excitation balance is shifted upwards since the bf cross
section of the ground state is small and since there simply are more
photons at the longer wavelengths of the excited levels' ionization
edges. This effect was found for B\,{\sc i} by Kiselman (1994b) and it
can be thought of as the opposite of the ``photon suction'' discussed
by Bruls, Rutten, \& Shchukina (1992) for Na\,{\sc i} and K\,{\sc i}.
Quenching the pumping effect of the ultraviolet resonance line at
323\,nm by treating it as a fixed rate given by $J_\nu = B_\nu$ has a
very small impact ({\bf 323~nm~B}). This specific pumping is unimportant for
the formation of the 671\,nm line. When all radiative transitions
except the 671\,nm line are treated in the same way ({\bf bf\&bb~B}),
about half of the \nlte\ effect remains. The remaining departures from
\lte\ are due to the resonance line itself.

The importance of 3D effects is assessed by treating radiative
transitions as fixed using the precomputed 3D radiation field.
Doing this to the photoionizations causes
insignificant changes ({\bf bf~3D}).
Treating all radiative transitions, except the
671\,nm line, in 3D ({\bf bf\&bb~3D}), gives an
anticipated spread in the $I_c - W$  diagram since the local $J_\nu$ is
influenced by neighboring points in the snapshot.

Finally ({\bf all~3D}), {\em all} radiative transitions, including the
671\,nm resonance lines, are treated as fixed in 3D, now also with the
application of a correction for line blanketing in the background
radiation fields. This is the most sophisticated simulation presented here, but
its shortcomings must be remembered -- the 3D radiation fields have
been calculated in \lte\ in a somewhat schematic way, all radiative
transitions have been treated as fixed, and the correction for line
blanketing is schematic. The $I_c - W$ plot shows a rather flat slope
and a significantly larger spread than in the 1.5D case.
Apparently, effects of horizontal photon transfer are of some importance through
the line transitions. However, the difference in intensity-weighted
mean equivalent width between the 1.5D and the 3D case is only 3\,\%.

Experiments with higher abundances show that the line behavior
described here is
qualitatively the same at a ten times higher abundance (2.1), but at
an abundance of 3.1 we leave the regime of weak-line behavior and the typical
effects of strong resonance lines (Carlsson et al. 1994) set in.

\section{Conclusions}
Line formation in granulation is not well approximated with simple
``cold'' and ``hot'' regions or streams. The strongly varying vertical
photospheric temperature gradient, the typical inversion of the temperature
pattern above a certain height, and the presence of inclined
thermal inhomogeneities complicate the situation to the extent that
conclusions regarding spectral lines
must be based on realistic granulation simulations.
A case has been made that such simulations together with spatially
resolved solar spectroscopy can be used as diagnostics of
\nlte\ effects. Another paper (Kiselman 1997) will follow up on
this and compare simulations with solar observations of \lii.

The current calculations do not show evidence for any
large \nlte\ effects in quiet
granulation that would seriously affect lithium abundance
determinations for solar-like stars. This is not to say that effects
of granulation are altogether
unimportant -- more comprehensive studies are needed to quantify
``granulation abundance corrections'' corresponding to the commonly
used \nlte\ abundance corrections. One may also
speculate that 3D \nlte\ effects could be important in
regions of enhanced magnetic activity and thus be relevant to problems
such as the apparent lithium-abundance spread in the Pleiades
(cf. Stuik, Bruls, \& Rutten 1997).

To draw conclusions about very metal-poor solar-type stars and the \nlte\
effect proposed by Kurucz (1995) would be to
extrapolate -- the nature of these stars' surface
inhomogeneities is unknown but probably different from solar
(Allende~Prieto et al. 1995). One can note, however, that bound-bound
processes may be important for the statistical
equilibrium of Li there as well as in the solar case
studied here. Wherever line transitions are the dominating drivers of
departures from \lte, any 3D
effects are likely to be milder than if bound-free processes govern
the \nlte\ behavior, because the contrast of granulation or other
thermal inhomogeneity is lower at
the longer wavelengths of the \lii\ spectral lines than in the blue
and ultraviolet continua.

\acknowledgements
I thank Rob Rutten for discussions, comments, and manuscript sharing.
Mats Carlsson kindly provided the lithium-atom model.
G{\"o}ran Scharmer is thanked for comments on the manuscript and
Martin Asplund for discussions.

\newpage

\begin{figure}
\caption{ \lii\ 671\,nm behavior in a 3D hydrodynamic granulation snapshot. Upper row: \lte\ results. Lower row: 1.5D \nlte\
 results. Left: Equivalent width as function of continuum intensity.
 Middle and Right: Line source functions and total
 line-center opacities for six selected columns
 (A-F). Continuum intensities and source functions are normalized on
 the mean continuum intensity for the whole snapshot. Line-center
 opacities are normalized with the local continuum opacity. \label{fig_1.5D_33}} 
\end{figure}
\begin{figure}
\caption{\lii\ 671\,nm, results for 1D
 {\sc osmarcs} models with a range of $T_{\rm eff}$'s, same kinds of
 plots as in Fig.~\ref{fig_1.5D_33}. The continuum
 intensities are normalized to the model representing the Sun
 ($T_{\rm eff} = 5780\,{\rm K}$).  \label{fig_osmarcs_33} }
\end{figure}
\begin{figure}
\caption{The impact of various radiative-transfer treatments on
\lii\,671\,nm equivalent widths and their dependence on continuum
intensity. Explanations are given in text. $<W>$ is the
intensity-weighted mean equivalent width. \label{fig_icw22} }
\end{figure}

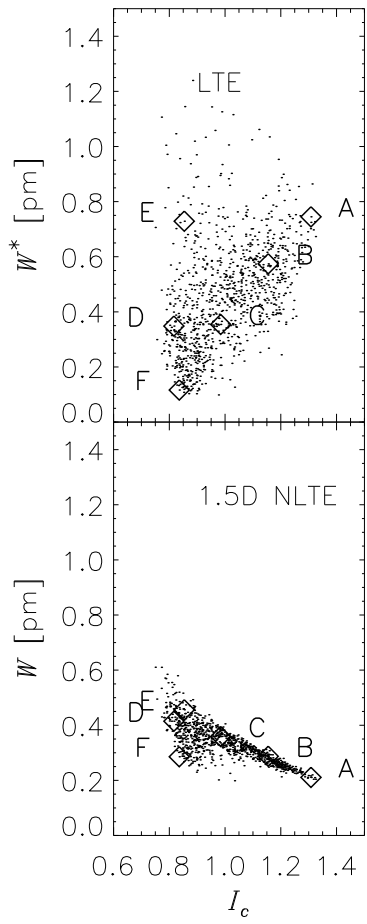
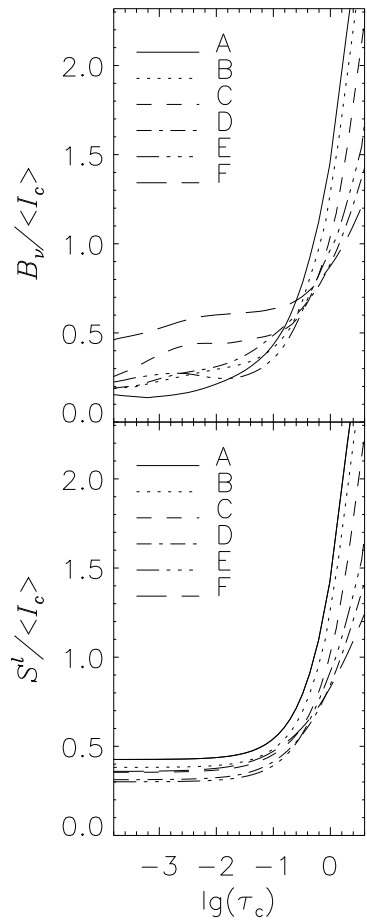
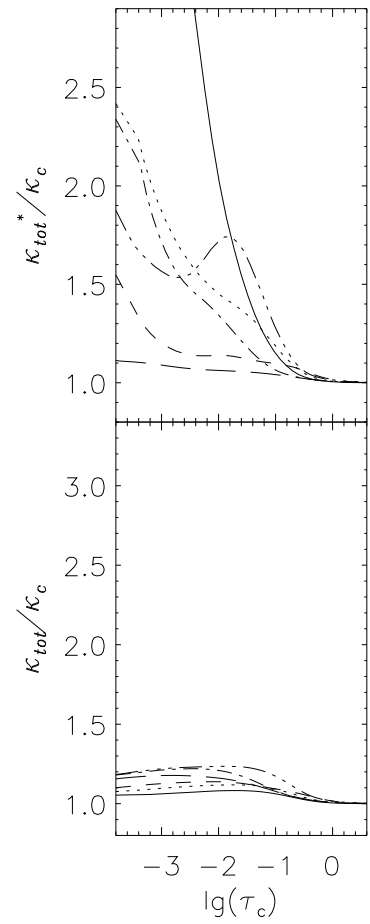
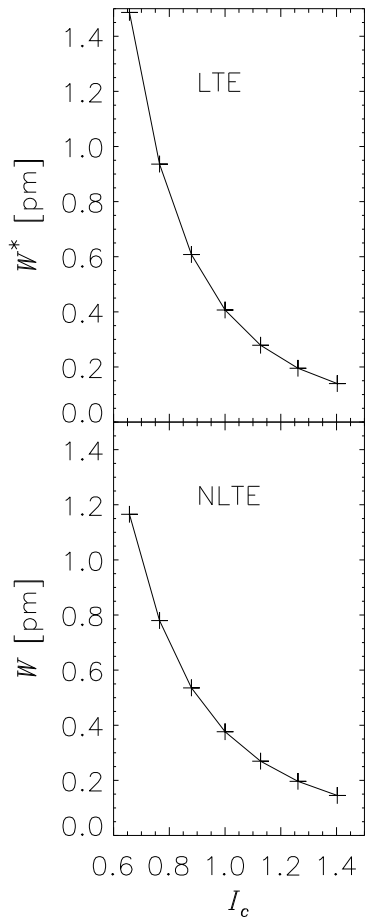
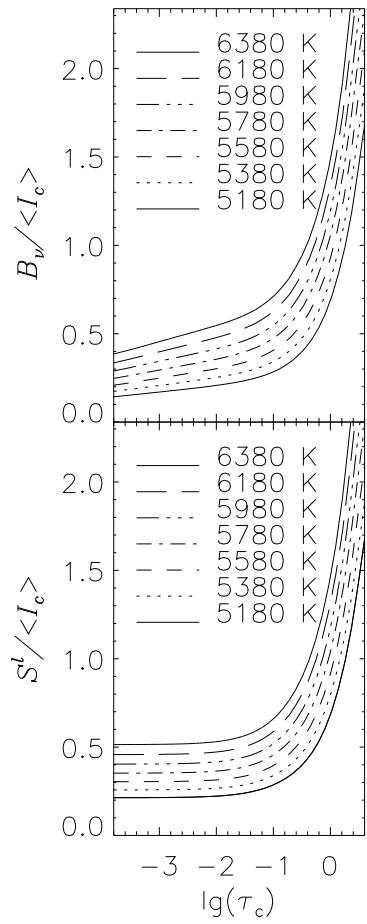
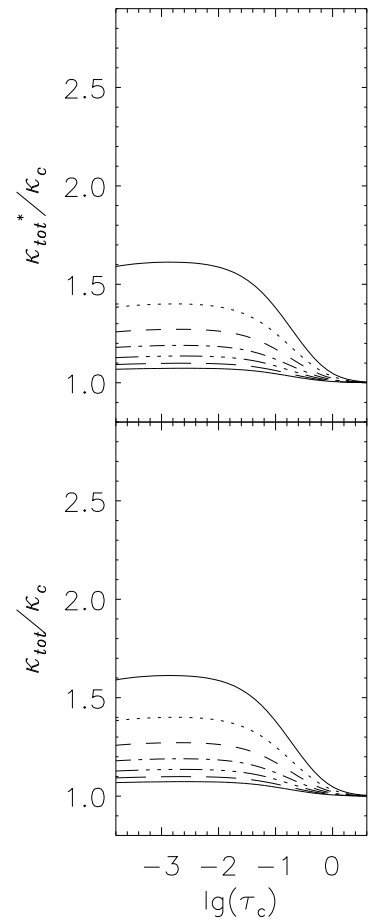
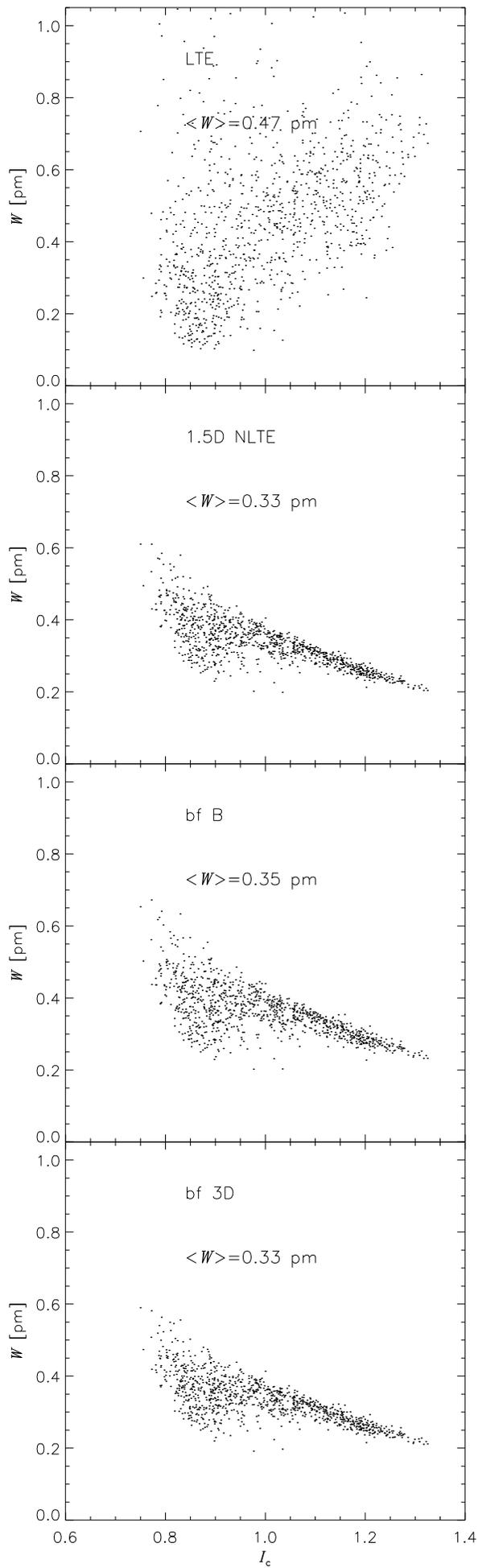
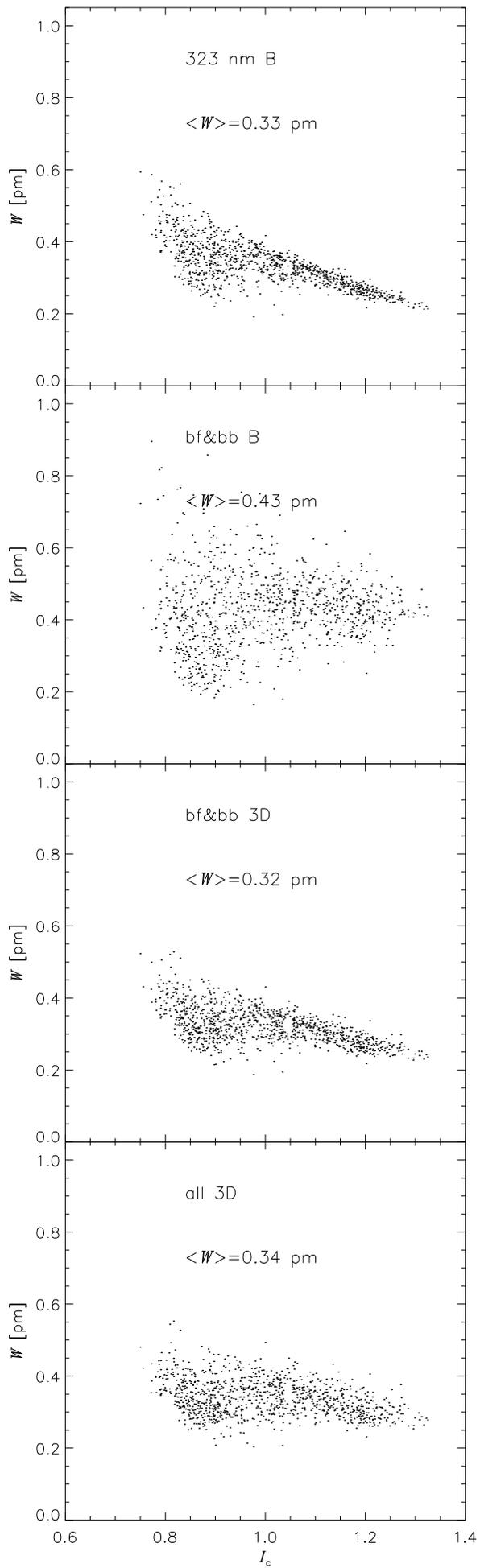
\end{document}